\documentstyle[12pt]{article}
\begin{document}
\begin{center}

{\large
{Quantum Field Theory for Orthofermions and Orthobosons}
}

\vskip 1.2cm

{\bf A. K. Mishra $^{*\dagger}$ and G. Rajasekaran $^{\dagger}$ }

\vskip 0.5cm
{\it {$^*$ Max-Planck Institute for Physics of Complex Systems,  
Nothnitzer Str. 38, D-01187 Dresden, Germany, and  \\
$^{\dagger}$ Institute of Mathematical Sciences,
CIT Campus, Madras - 600 113, India\\
e-mail: mishra@imsc.ernet.in; graj@imsc.ernet.in}}

\end{center}

\vskip 1.2cm

\baselineskip=18pt
\centerline{\bf Abstract}

\vskip 0.5cm

\noindent
Orthofermi statistics is characterized by  an exclusion principle
which is more ``exclusive'' than Pauli's exclusion principle: an
orbital state shall not contain more than one particle, no matter
what the spin direction is. The wavefunction is antisymmetric in
spatial indices alone with arbitrary symmetry in the spin indices.
Orthobose statistics is corresponding Bose analog: the wavefunction 
is symmetric in spatial indices, with arbitrary symmetry in spin
indices. We construct the quantum field theory of particles obeying
these new kinds of quantum statistics. Non-relativistic as well as
relativistic quantum field theories with interactions are considered.

\newpage

\baselineskip=20pt

\section*{1. Introduction}
The constraint that an orbital state shall not contain more than one 
particle irrespective of their spin directions has led us to the formulation 
of a new family of quantum statistics, namely, orthostatistics [1]. 
These statistics are described through the algebra

$$
c_{k\alpha} c^\dagger_{p\beta} \pm \delta_{\alpha \beta} \sum_\gamma c^\dagger_{p\gamma}
c_{k\gamma} \ = \ \delta_{kp} \delta_{\alpha \beta} 
\eqno(1)
$$
$$
c_{k\alpha} c_{p\beta} \pm c_{p\alpha} c_{k\beta} \ = \ 0 
\eqno(2)
$$ 
where
$c_{k\alpha}$ and $c^\dagger_{p\beta}$ are the annihilation and creation
operators of particles with momenta $k$ and $p$ and spins $\alpha$ and
$\beta$ respectively. The  positive sign corresponds to the orthofermi
and negative sign is for the orthobose statistics.

Greenberg has earlier constructed infinite statistics wherein 
wavefunctions with arbitrary symmetries are allowed [2,3]. Infinite statistics
also follows from Eq.(1) if the spatial indices $k$ and $p$
are suppressed [4]. On the other hand, if spin indices are suppressed,
canonical commutation relations for boson and anticommutaors for fermion are obtained.
Thus orthostatistics  describe a generalized class of statistics wherein
different indices exhibit different and uncorrelated symmetry properties. 

One of the important property of orthofermions is that only one particle can
be accommodated among the set of states $\{k \alpha \}$, irrespective of the
range of $\alpha$. This contrast with the occupancy status of parafermions.
In parafermi statistics of order $n$, utmost $n$ parafermions can
occupy the same state [5].  

Until now, it has been possible to construct a local relativistic
quantum field theory (LRQFT) only for  
parafermions and parabosons, of which fermion and boson are specific
examples [6]. 
No such formulation is possible for infinite statistics [2,3]. Since the
orthofermions (orthobosons) have properties common with fermions (bosons)
as well as with the particles obeying infinite statistics, we are motivated
to examine whether a LRQFT for orthoparticles can be constructed?
We show here that if the second index $\alpha$ in $c_{k\alpha}$ is
reinterpreted as a new degree of freedom which can be excited at some
higher energy scale yet to be probed experimentally, it is possible
to construct  LRQFT for orthofermions [7]. However, the relativistic quantum
field theory for the orhobosons remains nonlocal. 

The nonrelativistic
quantum field theory for the both kinds of particles can be constructed
even when $\alpha$ denotes the usual kinematic spin variable. This is
considered in Sec.2.
Formulations
of Dirac field as an orthofermi field and Klein-Gordon field as an orthobose
field, provided  $\alpha$ characterizes the new degree of freedom, are 
given in sections 3 and 4, respectively. Interactions are considered in Sec.5, 
followed by a section on summary and conclusions.

\section*{2. Nonrelativistic Quantum Field theory}
The orthoparticles obey  quantization rules different from the canonical
commutators and anticommutators. The Poisson brackets involving canonical
coordinate $\psi$ and canonical momentum $\pi$, that is, $[\psi,\pi]_{PB}$ and 
$[\psi,\psi]_{PB}$ are now replaced by the new quantum brackets corresponding
to Eqs.(1) and (2), or equivalently

$$
\psi_a({\bf x},0) \psi^\dagger_b({\bf y},0) \pm \delta_{ab} \sum_c \psi^\dagger_c ({\bf y},0) \psi_c({\bf x},0) \ = \
\delta_{ab} \delta^3({\bf x-y}) \
\eqno(3)
$$
$$
\psi_a({\bf x},0) \psi_b ({\bf y},0) \pm \psi_a({\bf y},0) \psi_b({\bf x},0) \ = \ 0
\eqno(4)
$$

In spite of different quantization procedure, the corresponding   
nonrelativistic field theory can be 
developed in a consistent manner. Consider the nonrelativistic field 
$\psi_a({\bf{x}},t)$ satisfying the Schrodinger equation

$$
\left( i \frac{\partial}{\partial t} + \frac{1}{2m} \bigtriangledown^2\right) \psi_a({\bf x},t) \ =
\ 0
\eqno(5)
$$ 
The fourier expansion for $\psi_a$ is

$$
\psi_a({\bf x},t) \ = \ \sum_{{\bf k},\alpha} c_{{\bf k}\alpha} \xi^{(\alpha)}_{a}e^{{\bf ik \cdot x}-i\omega_k t} \,,
\eqno(6)
$$ 
where
$\xi^{(\alpha)}_{a}$ is the spin wavefunction for the spin-component $\alpha$,  
and $\omega_k  =   k^2/(2m) $ is the nonrelativistic energy.  

The Hamiltonian of the system is

$$
H \ = \ \sum_{{\bf k},\alpha} \omega_k c^\dagger_{{\bf k}\alpha} c_{{\bf k}\alpha} 
\eqno(7)
$$
From Eqs.(1) and (2) (with either sign), the following commutation relation can be derived

$$
\left[ c_{{\bf k}\alpha}, \sum_\gamma c^\dagger_{{\bf p}\gamma} c_{{\bf p}\gamma}\right] \ = \ \delta_{{\bf kp}}
c_{{\bf k}\alpha}
\eqno(8)
$$ 
Now it can be shown that the
Schrodinger Eq.(5) follows from the Heisenberg equation of motion for $\psi_a$
and the relation (8). 

The Schrodinger equation and the Hamiltonian can also be obtained from the
canonical formalism. In fact the Schrodinger Eq.(5) is the 
Euler-Lagrange equation of motion  for the usual Lagrange density

$$
{\cal L} \ = \ i \sum_a \psi^\ast_a \frac{\partial \psi_a}{\partial t} - \frac{1}{2m} \sum_a \mbox{\boldmath $
\bigtriangledown $} \psi^\ast_a \cdot \mbox{\boldmath $ \bigtriangledown $} \psi_a \
\eqno(9)
$$
The Legendre transformation gives the corresponding total Hamiltonian as 

$$
H \ = \ \int d^3 x {\cal H} \ = \  \frac{1}{2m} \int d^3 x \sum_a \psi^\ast_a
\bigtriangledown^2 \psi_a\ 
\eqno(10)
$$ 
Substitution of Eq.(6) into Eq.(10) leads to the Hamiltonian
given by Eq.(7).  

Thus the analysis given in this section shows that our quantization is  
consistent, and a nonrelativistic quantum field theory based on a
canonical Lagrangian formalism
can be constructed for the orthoparticles.

\section*{3. Relativistic Quantum field theory for Orthofermions}
We consider the Dirac Hamiltonin
$$
H \ = \ \int \psi^\dagger (\mbox {\boldmath $\alpha$} \cdot {\bf  p} + \beta m) \psi d^3 x 
\eqno(11)
$$ 
and the usual four component Dirac  field $\psi_a({\bf x},t)$ ($ 
a $  = 1,2,3,4). 
Using the expansion

$$
\psi_a({\bf x},t) \ = \ \sum_{\bf k} \sum^4_{\alpha =1} \left( \frac{m}{E_k}\right)^{1/2} c_{{\bf k}\alpha}
u^{(\alpha)}_a ({\bf k}) e^{i{\bf k\cdot x}-iE^{(\alpha)}_k t}\,, 
\eqno(12)
$$ 
and the relation

$$
(\mbox{\boldmath $\alpha$} \cdot {\bf k} + \beta m)u^{(\alpha)} {\bf (k)} \ = \ E^{(\alpha)}_k 
u^{(\alpha)}({\bf k}) 
\eqno(13)
$$ 
in Eq.(11), the Hamiltonian can be written as 

$$  
H \ = \
 \sum_k E_k \left[ \sum_{\alpha = 1,2} c^\dagger_{{\bf k} \alpha} c_{{\bf k}\alpha} - \sum_{\alpha =3,4}
c^\dagger_{{\bf k} \alpha} c_{{\bf k}\alpha}\right] 
\eqno(14)
$$ 
$u^{(\alpha)}_a ({\bf k})$ in expansion (12) are Dirac spinors, $\alpha$ = 1 and 2
being the 
positive energy spinors $ (E_k^{(\alpha)} \equiv E_K  = +(m^2 + k^2)^2)$ , and $\alpha$ = 3 and 4  
being the negative energy spinors $(E_k^{(\alpha)} \equiv -E_k)$. 
\boldmath ${\bf \alpha}$ \unboldmath , $\beta$ are the Dirac matrices and 
${\bf p}$ =  -i \boldmath $ \bf {\bigtriangledown}$ \unboldmath.

If  $c_{{\bf k }\alpha}$ and $c^{\dagger}_{{\bf k}\alpha}$ satisfy the 
orthofermionic algebra
given in Eqs.(1) and (2), and the second index $\alpha$ for an orthfermion 
is identified as the Dirac index $\alpha$ going from 1 to 4, then neither
$ \sum_{\alpha = 1,2 } c^\dagger_{{\bf k}\alpha }
c_{{\bf k}\alpha} $ nor  
$ \sum_{\alpha = 3,4 } c^\dagger_{{\bf k}\alpha }
c_{{\bf k}\alpha} $ occurring in the Hamiltonian (Eq.(14)) are number operators.
It can be verified from Eq.(8) that only the sum  
$ \sum_{\alpha  = 1-4 } c^\dagger_{{\bf k}\alpha} 
c_{{\bf k}\alpha} $ is a number operator for particles of momenta ${\bf k}$. 
Therefore as such, the Hamiltonian in Eq.(14) can not be reexpressed as 
the sum of energies of particles, and our quantization procedure seems to
fail. Note that if $c^{\dagger}$ and $c$ in Eq.(14) denote creation and
annihilation operators for the usual fermion, the
Hamiltonian can be reexpressed
as the sum of energies of particles and antiparticles by introducing the Dirac 
vacuum state.

The above problem can be circumvented by introducing a new degree of freedom
indexed by $A, B, C, D...$. The  orthofermi algebra with the new 
indices is written as

$$
c_{{\bf k}\alpha A} c^\dagger_{{\bf p}\beta B} + \delta_{A B} \sum_D c^\dagger_{{\bf p}\beta D}
c_{{\bf k}\alpha D } \ = \ \delta_{{\bf kp}} \delta_{\alpha \beta} \delta_{A B}
\eqno(15)
$$
$$
c_{{\bf k}\alpha A} c_{{\bf p}\beta B} + c_{{\bf p}\beta A} c_{{\bf k}\alpha B} \ = \ 0 
\eqno(16)
$$ 
The Eq.(14) for the Hamiltonian now gets replaced as  

$$   
H \ = \
 \sum_k E_k \left[ \sum_{\alpha =1,2} \sum_A c^\dagger_{{\bf k}\alpha A} c_{{\bf k} \alpha A} -
\sum_{\alpha=3,4} \sum_A c^\dagger_{{\bf k}\alpha A} c_{{\bf k}\alpha A} \right]\
\eqno(17)
$$ 
It can be verified that 
$$
n_{{\bf k}\alpha} \ = \ \sum_A c^\dagger_{{\bf k}\alpha A} c_{{\bf k}\alpha A} 
\eqno(18)
$$
is the  number operator for particles of momentum ${\bf k}$ and Dirac 
index $\alpha$. Note that in orthostatistics, number operators
$n_{{\bf k} \alpha A}$ with all the indices specified, do not exist [8]. 
The Dirac vacuum is introduced as the  filled negative energy sea. The  vacuum 
state is infinitely degenerate since orthofermions of negative (as well as positive)
energy can have arbitrary index $A$. One way of lifting this degeneracy is to choose
the vacuum state  as the normalized sum of all the states with the new index taking the full
range of values.

The creation and annihilation operators for antiparticles are  defined as
$$ 
d_{{\bf k}\alpha A} \ = \ c^\dagger_{{\bf k}\alpha+2,A} \ ; \
d^\dagger_{{\bf k}\alpha A} \ = \ c_{{\bf k}\alpha+2,A}  \quad ; \quad \alpha = 1,2
\eqno(19)
$$
The number operator for antiparticles is

$$
\bar{n}_{{\bf k}\alpha} \ = \ 1 - n_{{\bf k}\alpha+2} = 1 - \sum_A c^\dagger_{{\bf k}\alpha +2,A} c_{{\bf k}\alpha
+2,A} ~~~ {\rm for} \ \alpha = 1,2 
\eqno(20)
$$ 
which can be rewritten as 
$
\bar{n}_{{\bf k}\alpha} \ = \ d^\dagger_{{\bf k}\alpha A} d_{{\bf k}\alpha A}$,  for any  $A$.
The $A$-independence of the product
$ d^\dagger_{{\bf k}\alpha A} d_{{\bf k}\alpha A} $ follows from the Eqs.(15) 
and (19).
Using Eqs.(18) and (20), the Eq.(17) can be rewritten as

$$
H \ = \ \sum_k E_k \sum_{\alpha = 1,2} (n_{k\alpha} + \bar{n}_{k\alpha}) - \sum_k E_k
\sum_{\alpha=1,2} 1
\eqno(21)
$$ 
Note that no summation over $A$ appears in the last term  describing the 
vacuum energy. In orthofermi statistics, for each ${\bf k}$ and $\alpha$, there
is only one particle irrespective of the value of $A$. Subtracting the 
vacuum energy, we finally get the modified Hamiltonian

$$
\widetilde{H} \ = \ H - <0|H|0> \ = \ \sum_k E_k \sum_{\alpha=1,2} (n_{{\bf k}\alpha} +
\bar{n}_{{\bf k}\alpha}) 
\eqno(22)
$$ 
which is positive definite and is expressed as the sum of energies of particles
 and antiparticles. That the 
Heisenberg equation of motion for $\psi_{a A}({\bf x}, t)$ is consistent
with the Dirac equation 

$$
i \frac{\partial \psi_{a A}}{\partial t} \ = \ (\mbox {\boldmath $\alpha$} \cdot {\bf p} + \beta m) \psi_{a A}  
\eqno(23)
$$ 
and the algebra of $c$ and $c^{\dagger}$ (Eq.(15),
can be easily verified.
The Hamiltonian (Eq.(17)) can also be 
derived from the Lagrangian density

$$
{\cal L} \ = \sum_{a A} \ i\psi_{a A}^\dagger \frac{\partial \psi_{a A}}{\partial t}  - \sum_{a A}
\psi_{a A}^\dagger(-i \mbox{\boldmath $\alpha \cdot \bigtriangledown $}+ \beta m)\psi_{a A} 
\eqno(24)
$$ 
Thus the consistency of quantization rule as given by Eqs.(15,16), or 
equivalently by the relations

$$
\psi_{aA}({\bf x},0) \psi^\dagger_{bB}({\bf y},0) + \delta_{AB} \sum_D \psi^\dagger_{bD}({\bf y},0)
\psi_{aD}({\bf x},0) = \delta_{AB} \delta_{ab} \delta^3({\bf x} - {\bf y}) 
\eqno(25)
$$
$$
\psi_{aA}({\bf x},0) \psi_{bB}({\bf y},0) + \psi_{bA}({\bf y},0) \psi_{aB}({\bf x} ,0) = 0\,. 
\eqno(26)
$$
is established.

In spite of the modified structure of the basic commutation relations
in Eqs.(25) and (26), bilinear observables such as the current density
$j_{\mu}({\bf x}, t)$ at two different points commute for space like separation, thus satisfying
the microcausality condition. We define the current density four-vector
$
j_\mu \ = \ \sum_A \psi^\dagger_A \gamma_0 \gamma_\mu \psi_A
$
where $
\gamma_0 \ = \  \beta $ and  \boldmath $ \bf { \gamma}$ \unboldmath  = $\gamma_0$ \boldmath $\bf {\alpha}$ \unboldmath. 
It is straightforward to show that

$$
[j_\mu({\bf x},0), j_\nu({\bf y},0)] \ = \ 0 \ {\rm for} \ x\ne y
\eqno(27)
$$
Because of the relativistic invariance, commutativity of
$j_\mu({\bf x},t)$ and $ j_\nu({\bf y},t^{\prime})$ for arbitrary space like separations 
follows from the above relation.

\section*{4. Scalar Field as an orthobose Field}
The Eqs.(15) and (16) with the positive signs being replaced by negative signs 
define the algebra for orthobosons. We consider a real (Hermitian) scalar
field

$$  
\phi_A({\bf x},t) \ = \ \phi^\dagger_A({\bf x},t) 
\ = \ \sum_k \frac{1}{(2\omega_k)^{1/2}} (c_{{\bf k}A} e^{i{\bf k \cdot x}-i\omega_k t} +
c^\dagger_{{\bf k}A} e^{-i{\bf k \cdot x}+i\omega_k t)}
\eqno(28)
$$
$
\omega_k \ = \ +( k + m^2)^{1/2}  
$. The Hamiltonian is taken as  

$$ 
H \ = \ \frac{1}{2} N \int \sum_A (\mbox{\boldmath $\bigtriangledown $} \phi_A \cdot \mbox{\boldmath $
\bigtriangledown $}\phi_A + m^2 \phi_A \phi_A + \dot{\phi_A} \dot{\phi}_A) d^3 x
\ = \ \sum_{\bf k} \omega_k n_{{\bf k}}
\eqno(29)
$$ 
where $N$ denotes the normal ordering operator. The number operator for definite
momentum $ {\bf k}$ is $n_{{\bf k} A} = \sum_A c_{{\bf k}A}^{\dagger} c_{{\bf k} A}$. If the Hamiltonian had been
defined without normal ordering, $\omega_k$ in Eq.(29) would have been 
scaled by a term $(1 + n)/2$, where $n$ is the range of the $A$. The operator
$N$ removes this unacceptable scaling term along with the zero point
energy.

The consistency of the quantization can  be established by showing that
Klein-Gordon equation for $\phi$ follows from the Heisenberg equation of 
motion and orthobose algebra. The Hamiltonian density $ {\cal H}(x)$ in 
Eq.(29) can also be derived by starting with the Lagrangian density

$$
{\cal L} \ = \ \sum_A \frac{1}{2} \left\{\frac{\partial \phi_A}{\partial t}
\frac{\partial \phi_A}{\partial t} - \mbox {\boldmath $\bigtriangledown $} \phi_A \cdot \mbox{\boldmath $
\bigtriangledown $}\phi_A - m^2 \phi_A \phi_A \right\} 
\eqno(30)
$$ 
and employing the Legendre transformation.

However, if we take the Hamiltonian density as a local operator, it can be shown
that ${\cal H}({\bf x},0)$ and ${\cal H}({\bf y}, 0)$ do not commute
for ${\bf x} \ne {\bf y}$. The commutation relations among $\phi^S$
needed for this purpose can be derived using the Eq.(28) and orthobose algebra.
Thus the relativistic field theory for orthobose does  not satisfy the 
microcausality.

\section*{5. Interactions}
The interactions can be introduced in the relativistic theory in
usual way. A brief comment about the nature of new degree of freedom, which 
has been left unspecified so far, will help us in formulating appropriate
interactions. We may consider that the new degree of freedom is also 
described by a compact Lie group symmetry such as $SU(n)$ or $SO(n)$ etc.,
so that the field $\psi_A$ and $\phi_A$ form representations of Lie
algebra with index $A$ labelling the components of multiplet as in the 
usual quantum field theory. Of course, the physical consequences in the 
present case will be different.

The interaction terms can be now constructed which are invariant under 
the Lie group transformations. For example, if $\psi_A$ is an orthofermi
field which is a multiplet under $SU(n)$, and $\phi$ is an ordinary Bose field
which is singlet under $SU(n)$, an interaction term is given as
$\sum_A \psi_A^{\dagger} \beta \psi_A \phi$. 

\section*{6. Summary and conclusions}
In contrast to infinite statistics, we have been able to construct
local relativistic quantum field theory for orthofermions. The two
factors responsible for this are: (i) the number operators
are bilinear in $c^{\dagger}$ and $c$, and (ii) in orthofield theory,
quadratic relation in $c$ exists.

However, our success has been achieved by introducing a new degree of freedom.
In fact we have circumvented the problem faced by infinite statistics 
by allowing the conventional degrees of freedom to be associated with
the fermionic (antisymmetry) or bosonic (symmetric) behaviour in the 
wave function and assigning the new property of infinite statistics  
(arbitrary symmetry) to the new degree of freedom.

It has been pointed out that the infinite statistics can not be
based on a canonical Lagrangian formalism [9]. In contrast, orthofield
theory is based on a canonical Lagrangian formalism for the
Schrodinger and Dirac fields. 

It is most remarkable that microcausality is satisfied for the orthofermi field.
Thus, we have shown that even with a modified quantization procedure, it is possible
to obtain a consistent quantum field theory, and thus the framework of 
quantum field theory has been enlarged.


\newpage

\baselineskip=12pt

\begin{thebibliography}{99}
\item Mishra A. K., and Rajasekaran G.,
{\it Pramana - J. Phys.} {\bf 36}, 537-555 (1991); {\bf 37}, 455 (E) (1991).
\bibitem{Greenberg90}Greenberg O. W., {\it Phys. Rev. Lett.} {\bf 64},
705-708 (1990).
\item Greenberg O. W., {\it Phys. Rev.} {\bf D43},
4111-4120 (1991).
\item Mishra A. K., and Rajasekaran G.,
{\it Phys. Lett. A} {\bf 188}, 210-214 (1994).
\item Green H. S., {\it Phys. Rev.} {\bf 90}, 270-273 (1953).
\item Fredenhagen K., {\it Commun. Math. Phys.} 
{\bf 79}, 141-151 (1981).
\item Mishra A. K., and Rajasekaran G.,
{\it Mod. Phys. Lett. A} {\bf 7}, 3425-3437 (1992); {\bf 11}, 1031 (E) (1996).
\item Mishra A. K., and Rajasekaran G.,
{\it Pramana - J. Phys.} {\bf 45}, 91-139 (1995); hep-th/9605204.
\item Chaturvedi S., Kapoor A. K., Sandhya R.,
Srinivasan V., and Simon R., {\it Phys. Rev.} {\bf A43}, 4555-4557 (1991). 
\end{thebibliography}
\end{document}